\documentclass[conference]{IEEEtran}
\ifCLASSINFOpdf
\else
\fi
\usepackage{graphicx}
\usepackage{subfigure}
\begin{document}
%
\title{Fast Cycle Frequency Domain Feature Detection for Cognitive Radio Systems
}

\author{\IEEEauthorblockN{Shen Da\IEEEauthorrefmark{1},   Gan Xiaoying\IEEEauthorrefmark{1},    Chen Hsiao-Hwa\IEEEauthorrefmark{2}, Qian Liang\IEEEauthorrefmark{1}}
\IEEEauthorblockA{\IEEEauthorrefmark{1}Dept. of Electronic
Engineering, Shanghai Jiao Tong University, Shanghai, 200240,
China\\ Email: ganxiaoying@sjtu.edu.cn}
\IEEEauthorblockA{\IEEEauthorrefmark{2}Dept. of Engineering Science,
National Cheng Kung University, Tainan City, Taiwan} }


%


\maketitle
\renewcommand{\thefootnote}{\fnsymbol{footnote}}

\begin{abstract}
 In cognitive radio systems, one of the main requirements is to detect the presence of the primary users' transmission, especially in weak signal cases. Cyclostationary detection is always used to solve weak signal detection, however, the computational complexity prevents it from wide usage. In this paper, a fast cycle frequency domain feature detection algorithm has been proposed, in which only feature frequency with significant cyclic signature is considered for a certain modulation mode. Simulation results show that the proposed algorithm has remarkable performance gain than energy detection when supporting real-time detection with low computational
 complexity.
\end{abstract}


%
\IEEEpeerreviewmaketitle

\section{Introduction}
The remarkable growth of wireless services over the last decade demonstrates the vast and increasing demand for radio spectrum. However, the spectrum resource is limited and most has been licensed exclusively to users which can work within a limited frequency band. Recent study [1] shows that the actual licensed spectrum is largely unoccupied most of the time. Thus, cognitive radio (CR) has been proposed to solve this problem [2], [3]. By sensing and adapting to the environment, CR users are able to fill in spectrum holes and serve its users without causing harmful interference to the licensed user. Therefore, the CR system requires spectrum sensing technique that detects the unoccupied spectrum band as quickly and accurately as possible for its implementation.\\
\indent Various spectrum sensing techniques have been presented£¬including matched filter, energy detection and cyclostationary detection [4-7]. Matched filter requires perfect knowledge of the primary users' signal features. Energy detection method is sensitive to noise and interference level. As an alternative, cyclostationary detection has a good performance in low SNR scenarios, but it costs a large amount of computational capacity, which makes it not suitable for real-time detection [8]. In this paper, a fast cycle frequency domain feature detection algorithm has been proposed, in which only feature frequency with significant cyclic signature is considered for a certain modulation mode. The detection performance and computational complexity for the proposed algorithm are analyzed and compared with those of energy detection.\\
\indent The rest of this paper is organized as follows: The system
model under consideration is discussed in Section II. Section III
gives a brief overview of cyclostationary spectrum analysis. Section
IV introduces the proposed algorithm. Simulation results and
performance comparison with energy detection are given in Section
IV. Finally, conclusions are drawn in Section V.

\section{ SYSTEM MODEL}
The spectrum sensing problem can be modeled as hypothesis testing.
It is equivalent to distinguishing between the following two
hypotheses:

\begin{equation}
\left\{ \begin{array}{l}
 H_0: y(t) = n(t) \\
 H_1: y(t) = x(t) + n(t) \\
 \end{array} \right.
\end{equation}\\
$y(t)$,$x(t)$and $n(t)$ denote the received signal, the primary
user's transmit signal, and the noise, respectively. $H_1$ and $H_0$
represent the hypothesis that the primary user is active or
inactive. Due to the existence of noise, a certain threshold
$\lambda _{{\rm{th}}} $ should be set to decide whether a primary
user is active or not. Probability of detection ($P_d$) and false
alarm ($P_f$) are defined to evaluate the detection performance:

\begin{equation}
\left\{ \begin{array}{l}
 P_d  = p({\rm{y}}_i  > \lambda _{{\rm{th}}} \left| {H_1 } \right.) \\
 P_f  = p(y_i  > \lambda _{{\rm{th}}} \left| {H_0 } \right.) \\
 \end{array} \right.
\end{equation}
The goal of detection is to maximize the $P_d$ while maintain a
given $P_f$.

\section{CYCLOSTATIONARY SPECTRUM ANAYSIS}
The cyclic autocorrelation of a complex-valued time series $x(t)$ is
defined by [7]:

\begin{equation}
R_x^\alpha  \left( \tau  \right) \buildrel \Delta \over = \mathop
{\lim }\limits_{T \to \infty } \frac{1}{T}\int\limits_{ -
\frac{T}{2}}^{\frac{T}{2}} {x\left( {t + \frac{\tau }{2}} \right)}
x^* \left( {t - \frac{\tau }{2}} \right)e^{ - i2\pi \alpha t} dt
\end{equation}
where $ \alpha $ is referred to as the cycle frequency.\\
\indent The Spectral Correlation Density (SCD) of $x(t)$, which is
also known as the cyclic spectrum, can be obtained by Fourier
transforming the cyclic autocorrelation

\begin{equation}
S_x^\alpha  \left( f \right) = F\left\{ {R_x^\alpha  \left( \tau
\right)} \right\} = \int\limits_{ - \infty }^\infty  {R_x^\alpha
\left( \tau  \right)} e^{ - i2\pi f\tau } d\tau
\end{equation}\\
\indent If $x(t)$ is Additive White Gaussian Noise (AWGN), then when
$ \alpha  \ne 0 $,$ S_x^\alpha  (f) = 0 $.\\
\indent The frequency-smoothing method is adopted to estimate SCD,
and the discrete formulation can be expressed as [9] [10]:

\begin{equation}
\begin{array}{l}
 \tilde S_x^\alpha  \left[ l \right] = \frac{1}{{\left( {N - 1} \right)T_s }}* \\
 \frac{1}{L}\sum\limits_{v = \frac{{ - \left( {L - 1} \right)}}{2}}^{v = \frac{{L - 1}}{2}} {X\left( {l + \frac{\alpha }{{2F_s }} + v} \right)X^*\left( {l - \frac{\alpha }{{2F_s }} + v} \right)W\left( v \right)}  \\
 \end{array}
\end{equation}
where
\begin{equation}
X\left[ v \right] = \sum\limits_{k = 0}^{N - 1} {x\left[ k \right]}
e^{ - i2\pi vk/N}
\end{equation}
$N$ is the number of input signal samples, $ x\left[ k \right] =
x\left( {kT_{\rm{s}} } \right)$, $T_{\rm{s}}$ is the time-sampling
increment, $ F_s  = {\raise0.7ex\hbox{$1$} \!\mathord{\left/
 {\vphantom {1 {NT_s }}}\right.\kern-\nulldelimiterspace}
\!\lower0.7ex\hbox{${NT_s }$}}$ is the frequency-sampling increment,
$ l = \left\lfloor {{\raise0.7ex\hbox{$f$} \!\mathord{\left/
 {\vphantom {f {F_s }}}\right.\kern-\nulldelimiterspace}
\!\lower0.7ex\hbox{${F_s }$}}} \right\rfloor $ is the discrete value
of the frequency, and $W\left( v \right)$ is the frequency-smoothing
function centered at $v = 0$ of width $L$ .

\section{ Cycle Frequency Domain Feature Detection Algorithm}
The Spectral Correlation Density (SCD) of a signal is a
cross-correlation function between frequency components separated by
$f + \frac{\alpha }{2} $and $f - \frac{\alpha }{2}$, we mapped it
from $f \sim \alpha$ square to $ \alpha $ axis through the following
expression
\begin{equation}
I_x \left( \alpha  \right) = \mathop {\max }\limits_f |S_x^\alpha
\left( f \right)|
\end{equation}\\
\indent And $ I_x \left( \alpha  \right) $ can be regarded as the
SCD of a signal at the cyclic frequency domain. Different signal
modulations will exhibit different cyclic features in the cyclic
frequency domain, as shown in Figure 1 and 2. AM and BPSK modulated
signals both have significant features at $ \alpha _0  = 2f_c $.\\

\begin{figure}[htbp]
\centering
\includegraphics[width=0.4\textwidth]{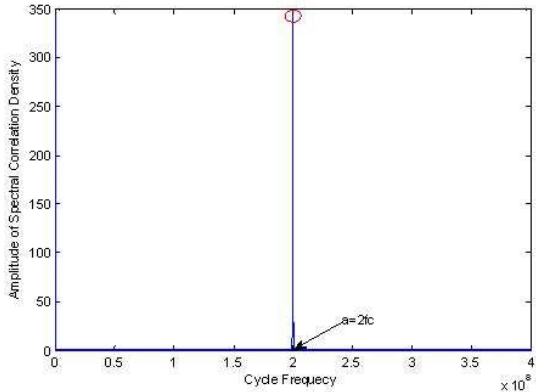}
\caption{Cycle Frequency Domain for AM signal}\label{fig_sim}
\end{figure}
\begin{figure}[htbp]
\centering
\includegraphics[width=0.4\textwidth]{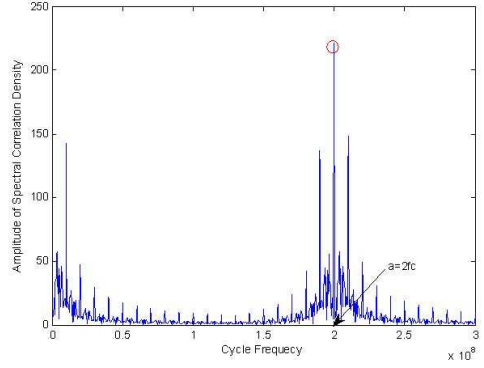}
\caption{Cycle Frequency Domain for BPSK signal}\label{fig_sim}
\end{figure}
\indent The cyclic features of a signal are decided by the
modulation types it employed. From the experiment results we found
that there exists a significant feature at $ \alpha _0  = 2f_c $ for
both analog and digital amplitude modulated signals, where the
carrier frequency is $f_c $. In the following section of this paper,
only SCD of the most significant features ($ \alpha _0 $) of a
modulated signal will be considered when performing detection.\\
\indent We assume that the modulation type and the carrier frequency
$ f_c $ of the signal are known, namely the cycle frequency $ \alpha
_0  = 2f_c $ is known to us. After SCD estimation at cycle frequency
$ \alpha _0  $ , the system detection model (1) changes into the
following form:
\begin{equation}
\left\{ \begin{array}{l}
 H_0 :\tilde S_y^{\alpha _0 } (l) = \tilde S_n^{\alpha _0 } (l) \\
 H_1 :\tilde S_y^{\alpha _0 } (l) = \tilde S_x^{\alpha _0 } (l) + \tilde S_n^{\alpha _0 } (l) \\
 \end{array} \right.
\end{equation}\\
\indent The cyclostationary spectrum sensing metric of the received
signal is obtained by
\begin{equation}
M = \mathop {\max }\limits_l \left| {\tilde S_y^{\alpha _0 } \left(
l \right)} \right|
\end{equation}\\
\indent A. When the primary user is inactive ( $H_0 $ hypothesis),
the sensing metric becomes

\begin{equation}
M_0  = \mathop {\max }\limits_l \left| {\tilde S_y^{\alpha _0 }
\left( l \right)} \right| = \mathop {\max }\limits_l |\tilde
S_n^{\alpha _0 } \left( l \right)|
\end{equation}
Theoretically, $n(t)$ is AWGN, so when $ \alpha _0  \ne 0 $, $\tilde
S_n^{\alpha _0 } (l) = 0 $, $ M_0  = 0 $. However, for Limited
length SCD, when $ \alpha _0  \ne 0 $, $ \tilde S_n^{\alpha _0 } (l)
\ne 0 $, $ M_0  \ne 0 $ [11].\\
\indent B. When the primary user is active ($H_1 $ hypothesis), the
sensing metric becomes

\begin{equation}
\begin{array}{l}
 M_1  = \mathop {\max }\limits_l |\tilde S_y^{\alpha _0 } (l)| \\
 {\rm{     }} = \mathop {\max }\limits_l |\tilde S_x^{\alpha _0 } (l) + \tilde S_n^{\alpha _0 } (l)| \\
 {\rm{     }} = \mathop {\max }\limits_l \left( {|\tilde S_x^{\alpha _0 } (l)| + |\tilde S_n^{\alpha _0 } (l)|} \right) \\
 {\rm{     }}\mathop { \approx \max }\limits_l |\tilde S_x^{\alpha _0 } (l)| + \mathop {\max }\limits_l |\tilde S_n^{\alpha _0 } (l)| \\
 {\rm{     }} = \mathop {\max }\limits_l |\tilde S_x^{\alpha _0 } (l)| + M_0  \\
 \end{array}
\end{equation}\\
\indent The decision on the presence of a primary user is simplified
to distinguish $M_0$ from $M_1$. when the sensing metric of the
receiving signal is M, namely decide the channel belongs to $H_0$ or
$H_1$.  A decision threshold $ \lambda _{th} $ is set to classify
the metric M:

\begin{equation}
\left\{ \begin{array}{l}
 M_0 (H_0 ):\begin{array}{*{20}c}
   {}  \\
\end{array}if\begin{array}{*{20}c}
   {M < \lambda _{th} }  \\
\end{array} \\
 M_1 (H_1 ):\begin{array}{*{20}c}
   {}  \\
\end{array}if\begin{array}{*{20}c}
   {M > \lambda _{th} }  \\
\end{array} \\
 \end{array} \right.
\end{equation}\\
\indent Probability of detection $P_d$  and probability of false
alarm $P_f$ then be formulated as

\begin{equation}
\left\{ {\begin{array}{*{20}c}
   {P_d  = \Pr (M > \lambda _{{\rm{th}}} |H_1 )}  \\
   {P_f  = \Pr (M > \lambda _{th} |H_0 )}  \\
\end{array}} \right.
\end{equation}\\
\indent The threshold $ \lambda _{th} $ can be selected for finding
an optimum balance between $P_d$ and $P_f$ .
\section{ SIMULATION RESULTS}
In this section, Monte Carlo simulation results are presented to
show the better detection performance for the proposed algorithm
when comparing with energy detection. Simulation parameters are
listed in TABLE 1.
\begin{table}[htbp]
\renewcommand{\arraystretch}{1.5}
\caption{Simulation parameters list} \label{table_example}
\centering
\begin{tabular}{|c|c|}
\hline
Parameter & Value\\
\hline
Modulation type & AM\\
\hline
Carrier frequency &  1 MHz\\
\hline
  Bandwidth & 10 KHz\\
\hline
  Sampling frequency & 3 MHz\\
  \hline
   Sampling time & 1.365ms\\
  \hline
  Channel & AWGN\\
  \hline
  Window type & hamming\\
  \hline
   Frequency smoothing length(L) & 1300\\
  \hline
  Sampled data length(N)& 4096\\
\hline
\end{tabular}
\end{table}\\
\indent Figure 3 shows the curves of receiver operating
characteristics (ROC) of the proposed method and energy detection,
it's obvious that the proposed method outperforms the energy
detection algorithm.\\
\indent Energy detection performs badly when SNR bellows -17db,
while proposed method still has a good performance under even -22db.
For example, when SNR equals to -22db and the probability of false
alarm ($P_f$) is 0.1, the probability of detection ($P_d$) of the
proposed method and energy detection equal to 0.99 and 0.18,
respectively; when $P_f=0.01$ , the two $P_d$ change to be 0.93 and
0.25, respectively. A cognitive radio operation in licensed TV bands
(IEEE 802.22 working group) defines "required SNR sensitivity" for
primary user signals to be: -22dB for DTV signals and -10 dB for
wireless microphones [12]. Therefore, the proposed method meets the
requirements well.
\begin{figure}[htbp]
\centering
\includegraphics[width=0.4\textwidth]{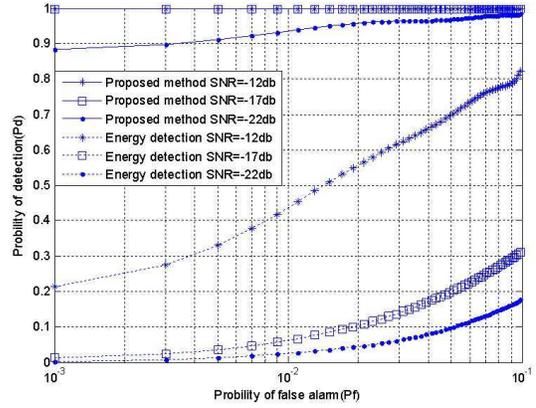}
\caption{Performance comparison between the proposed method and
energy detection}\label{fig_sim}
\end{figure}\\
\indent If we choose FFT to deal with the formulation (5) and use
the following expression to calculate the energy of a signal
($E_x$), the complexity of the two methods are shown in table 2.

\begin{equation}
E_x  = \sum\limits_{i = 1}^N {x\left( i \right)} x^* \left( i
\right)
\end{equation}
where $x(i)$ is the sample of a signal, N is the sample number.

\begin{table}[htbp]
\renewcommand{\arraystretch}{1.5}
\caption{Complexity comparison between the proposed method and
energy detection} \label{table_example} \centering
\begin{tabular}{|c|c|c|}
\hline
 & The proposed method & Energy detection\\
\hline Real multiply & $ 2N\log _2 N + 5L $ & $ 4N
$\\
\hline Real add & $ 3N\log _2 N + 3L $ & $ 3N
$\\
\hline
\end{tabular}
\end{table}
\indent From Table 2, we can see that the complexity of the proposed
method is approximately $ \log _2 N $ times of energy detection. The
main computation consumption is from FFT calculation in the method,
for a modern FFT chip this is not a big matter as if N is not too
big. So the more extra computation complexity of the proposed method
than energy detection is worthy, when considering the performance it
achieves. To achieve better detection performance, energy detection
demands much longer sampling time [13], which is not necessary for
the proposed method. Therefore, the proposed Fast Cycle Frequency
Domain Feature Detection algorithm support real-time detection as
well.

\section{Conclusion}
In this paper, a fast cycle frequency domain feature detection
algorithm is proposed, in which only feature frequency with
significant cyclic signature is calculated. Compared to the
traditional implementation of cyclostationary detection, computation
complexity for the proposed algorithm is greatly reduced under the
finite prior knowledge of modulation type and carrier frequency.
Simulation results show that the detection performance of the
proposed method outperforms that of energy detection. Therefore, the
proposed method is more suitable for real time spectrum sensing in
cognitive radio system.





%

\end{document}